\documentclass[twocolumn]{aastex61}

\usepackage{ulem}
\usepackage{color}
\usepackage{seqsplit}
\usepackage{hyperref}

\newcommand{\shao}{{Shanghai Astronomical Observatory, Chinese Academy of     
Sciences, Shanghai 200030, China}}

\submitjournal{AJ}

\begin{document}


\title{     {The localization of single pulse in VLBI observation}}
\author{Lei Liu}
\affiliation{\shao}

\author{Wu Jiang}
\affiliation{\shao}

\author{Weimin Zheng}
\affiliation{\shao}
\affiliation{Key Laboratory of Radio Astronomy, Chinese Academy of 
	Sciences, Nanjing 210008, China}
\affiliation{Shanghai Key Laboratory of Space Navigation and Positioning 	
	Techniques, Shanghai 200030, China}

\author{Zhen Yan}
\affiliation{\shao}

\author{Juan Zhang}
\affiliation{\shao}

\author{Maoli Ma}
\affiliation{\shao}

\author{Wentao Luo}
\affiliation{Kavli Institute for the Physics and Mathematics of the Universe 
(Kavli IPMU, WPI), University of Tokyo, Chiba 277-8582, Japan}

\correspondingauthor{Lei Liu}
\email{liulei@shao.ac.cn}


\begin{abstract}
     {In our previous work, we propose a cross spectrum based method to extract 
single 
pulse signals from RFI contaminated data, which is originated from geodetic 
VLBI postprocessing.
This method fully utilizes fringe phase information of the cross spectrum and 
hence maximizes the signal power. However, the localization has not been 
discussed in that work yet. As the continuation of that work, in this paper, we 
further study how to localize single pulses using astrometric solving method. 
Assuming that the burst is a point source, we derive the burst position by solving a 
set of linear equations given the relation between the residual delay and the 
offset to a priori position. We find that the 
single pulse localization results given by both astrometric solving and radio 
imaging are consistent within 3 $\sigma$ level. Therefore we claim that it is 
possible to derive the position of a single pulse with reasonable precision 
based on only 3 or even 2 baselines with 4 milliseconds integration. The combination of 
cross spectrum based detection and the localization proposed in this work then provide a 
thorough solution for searching single pulse in VLBI observation. According to 
our 
calculation, our pipeline gives comparable accuracy as radio imaging pipeline. 
Moreover,  
the computational cost of our pipeline is much smaller, which makes it more 
practical for FRB search in regular VLBI observation. The pipeline is now 
publicly available and we name it as ``VOLKS'', which is the acronym of ``VLBI 
Observation for frb Localization Keen Searcher''.}

\end{abstract}


\section{Introduction} \label{sec:intro}

The search of FRB \citep[Fast Radio Burst,][]{Lorimer2007} is now becoming an 
important 
topic in time domain astronomy.      {Their high precision localization is 
crucial in finding the possible background counterpart and finally explain the 
burst mechanism.       {Since the first discovery of FRB, so far only 
about 65 FRBs are found 
\citep{FRBCAT}.} In FRB searching, large 
single dish telescopes firstly play an important role \citep{Lorimer2007, 
Thornton2013, Ravi2015, 
Petroff2017, Bhandari2018}. However, the resolution of single dish telescope is 
of arcminute level, which is too large to isolate the transients
from background sources or associate them with possible counterparts 
\citep{Chatterjee2017}. In this case, interferometers with higher angular 
resolution provide another choice. To fully explore the performance of 
different types of interferometric instruments, several single pulse search 
methods are proposed \citep{Law2011}, including beam forming, radio imaging, 
etc.}

     {
Aperture arrays such as UTMOST \citep{Caleb2016} are dedicated to FRB 
search, ASKAP and CHIME \citep{CHIME17, CHIME18} take FRB search as one of 
their main	scientific goals. These arrays take the beam forming 
	approach\footnote{ASKAP antennas are equipped with phased array feed 
		(PAF): the whole focal plane is sampled and beams are formed 
		computationally 
		by combining signals from multiple PAF elements with complex 
		coefficients 
		(weights)\citep{ASKAP}. In this way ASKAP obtains good angular 
		resolution and 
		increases field of view simultaneously.}, in 
	which radio signals from multiple receivers are aligned in both time and 
	frequency 
	domain, and are then combined together to form multiple data beams to cover 
	large 
	searching area. After that these beams are searched for single pulses using 
	similar method for data from large single dish telescopes.
	Until now, UTMOST has successfully detected 4 FRB events \citep{Caleb2017, 
		Farah2018}.       {CHIME report detections of 13 FRBs at radio 
		frequencies as 
		low as 400~MHz, including one repeating burst \citep{CHIME, CHIME_rpt}.}
	\citet{Shannon2018} report the discovery 
	of 23 FRBs in a fly's-eye survey with ASKAP, which almost doubles the 
	number of 
	known events. Based on this sample, \citet{Macquart2018} derive a mean 
	spectral 
	index of -1.6.}

VLBI      {(Very Long Baseline Interferometry)} as the astronomical technique 
with the highest angular resolution \citep{Thompson2001}, is 
expected to provide high precision FRB localization. However, due to 
the relatively small FoV (Field of View), the search is usually carried out as 
commensal task 
in regular VLBI observations, e.g., V-FASTR \citep{VFASTR2011, Thompson2011} 
for VLBA, LOCATe for EVN \citep{LOCATe}. In these projects, the station auto 
spectrum is first dedispersed and searched for single pulses, then candidates 
from multiple stations are cross matched. This method is fast and 
easy to implement. However, it does not utilize the cross 
spectrum fringe phase information.      {According to our study in 
\citet{Liu2018b}, this potentially reduces its single pulse detection 
capability with RFI contaminated data.} In this case, cross spectrum based 
search methods are more 
suitable for VLBI observation. Among these methods, the most successful one is 
radio imaging, which detects single pulses in fast dumped images 
\citep{realfast}. 

In the rarest occasion of repeating FRB, localization can be in principle 
measured accurately. The position       {of the first discovered repeating burst FRB 
121102} is measured  \citep{Chatterjee2017, Marcote2017}      {with VLBI 
observation} and even the possible counterpart is identified in other bands \citep{Tendulkar2017, Bassa2017, 
Scholz2017}. In that work, the 
radio imaging search pipeline ``realfast'' plays a key role in 
detecting and localizing the burst in VLA data \citep{Chatterjee2017}. Other 
non-imaging methods also exist, e.g., the $uv$-fitting method and the 
bispectrum 
method which have been proposed and tested with the ``PoCo'' data 
\citep{Law2011, Law2012}.
However, due to some reasons, these methods are not widely deployed in current
search projects.

Although ``realfast'' has achieved great success      {for its detection and 
localization of repeating bursts of FRB 121102,} our calculation in 
Sec.~\ref{sec:computation} suggest that when it comes to VLBI observation with 
much longer baseline and therefore much higher angular resolution, to cover 
similar searching area      {as VLA antenna}, map size (pixel number along one 
side) becomes 
several orders of magnitude larger. The corresponding computational cost 
increases respectively, which makes it difficult to carry out FRB search in 
real VLBI observation. 

In \citet{Liu2018a}, we propose a geodetic VLBI based single pulse detection 
method. It 
takes the idea of geodetic VLBI fringe fitting that utilizes cross 
spectrum fringe phase information to maximize the signal power. Compared with 
auto spectrum based method, it is able to extract single pulses from highly RFI 
contaminated data \citep{Liu2018b}.      {As a continuation of that work, and 
to construct the whole single pulse search and localization pipeline,} in this 
paper, we further propose to localize single pulses in an astrometric solving 
approach: by assuming the burst is a point source, we may derive its accurate 
position      {by solving a set of linear equations based on the relation 
between the residual delay and the correction to a priori position.}
Compared with radio imaging, astrometric solving is much faster, and gives 
comparable accuracy. 

Our final goal for developing a      {complete single pulse search and 
localization} pipeline is to carry out 
FRB search in regular VLBI observations. According to our calculation 
\citep{Liu2018a}, by
assuming a reasonable event rate \citep{Keane2015} and appropriate spectral 
index and fluence index \citep{Caleb2017}, the FRB detection rate for VGOS 
\citep[VLBI2010 Global Observation System,][]{VGOS} antenna is 0.0076 events 
per sky 
per day. By assuming a 50\% observation efficiency, the expected detection rate 
is 1.387 events per year. Moreover, the number could double if we take a 
higher FRB event rate \citep{Champion2016}. Therefore, searching 
FRB and other transient events in regular VLBI observations are technically 
feasible and scientifically promising. 

      {For the real deployment of FRB search pipeline, we have to admit that 
there are still some problems that must be solved. One question which is often 
asked is whether the burst is detectable when it appears far from the phase 
center. Besides that, the performance of DM search with our cross spectrum 
based method has not been tested since the DM of current pulsar data set is too 
low. We present detailed discussions of these two issues in 
Sec.~\ref{sec:large_fov} and~\ref{sec:dm_search}, respectively.}

In this paper, we introduce the astrometric solving based single pulse 
localization method, compare its localization accuracy with radio imaging, and 
analyze the computational cost of our geodetic VLBI based pipeline and radio 
imaging pipeline.
This paper is organized as follows: In Sec.~\ref{sec:intro_methods}, we 
introduce the astrometric solving based localization method. In 
Sec.~\ref{sec:loc}, we present the single pulse localization result 
with a VLBI pulsar data set. In Sec.~\ref{sec:computation}, we analyze the 
computational cost of our pipeline and radio imaging. In  
Sec.~\ref{sec:discuss}, we discuss unresolved issues in current pipeline.  In 
Sec.~\ref{sec:conclusion}, we give our conclusion.

\section{The astrometric solving method for single pulse localization} 
\label{sec:intro_methods}

\begin{figure}
	\plotone{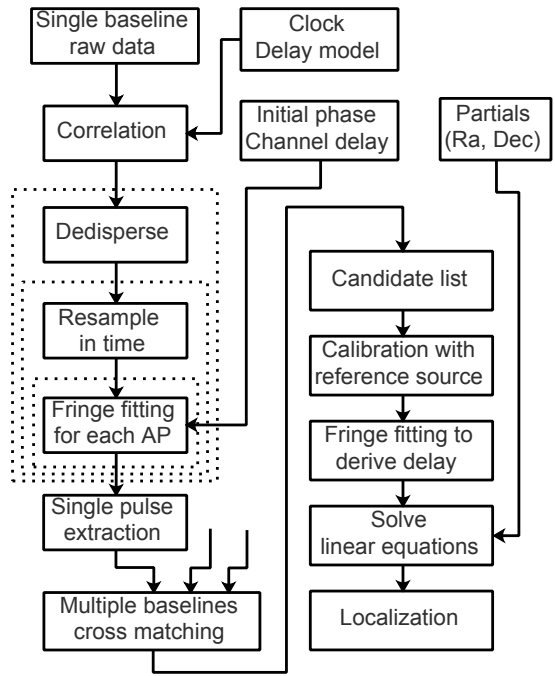}
	\caption{Demonstration of geodetic VLBI based single pulse search and 
	localization pipeline. Steps enclosed 
	by dotted rectangles represent iterations with corresponding 
	quantities.\label{fig:solving}}
\end{figure}

     {In this section, we introduce the astrometric solving method for single 
pulse 
localization. The localization part, together with the search part 
in \citet{Liu2018a}, build up the complete pipeline for single pulse search and 
localization in a geodetic VLBI approach.}

By assuming the burst is a point source, we are able to derive single pulse 
positions by solving a set of linear equations based on the relation between the
residual delay and the correction to a priori position. The idea is 
taken directly from solutions for earth orientation parameters (EOPs, usually 
include precession and nutation, polar motion, universal time), baseline and 
source position vectors, etc. in standard geodetic and astrometric VLBI 
measurement \citep{Thompson2001, Takahashi2000}. 
     {Actually there is already work that tries to obtain the high 
precision pulsar 
position in a geodetic VLBI approach: \citet{Sekido1999} obtain the position of 
B0329+54 and further derive its proper motion by first deriving the group delay 
with Japanese bandwidth synthesis 
software ``KOMB'' \citep{KOMB}, then carrying out analysis with CALC/SOLVE. 
However, their approach is still quite different from our work: they treat the 
pulsar as a normal radio source. Since B0329+54 is strong, they even do not 
use pulse gating. The total duration of pulsar observation that is used to 
derive position is as long as several hours. In contrast, in our work, we try 
to resolve and localize every individual single pulse with durations as short 
as 4 milliseconds. Actually we are the first that try to localize the single 
pulse in an astrometric solving approach.}

Concerning this work, since 
the pulsar is observed in phase reference mode, most of geodetic and 
atmospheric effects can be removed by phase reference calibration. What we need 
to estimate is just the correction to a priori position for every individual 
single pulse. Assuming the single pulse cross spectrum has been phase 
reference calibrated, for each baseline, the linear relation between the 
residual delay and the correction to a priori position can be expressed as:
\begin{equation}
\tau = \frac{\partial\tau}{\partial\alpha}\Delta\alpha + 
\frac{\partial\tau}{\partial\delta}\Delta\delta,
\end{equation}
where $\tau$ is the residual delay of this baseline, 
$\frac{\partial\tau}{\partial\alpha}$ and 
$\frac{\partial\tau}{\partial\delta}$ are partial derivatives of delay by
Ra and Dec, $\Delta\alpha$ and $\Delta\delta$ are corrections to a priori 
position. The residual delay can be derived 
by fitting the fringe phase. Two partial derivatives of delay by the source 
position are given by VLBI delay model, and can be obtained from popular 
model calculation programs, e.g., CALC. The above equations are solvable 
with two or more baselines. The least square solutions that takes 
the uncertainties of residual delay into account is described in 
Appendix~\ref{append:lsq}.

Fig.~\ref{fig:solving} demonstrates the whole geodetic VLBI based 
single pulse search pipeline. In this pipeline, the search and localization are 
two independent steps. In the first step, 
single pulse candidates are extracted by cross spectrum fringe fitting. Cross 
spectrum that takes the single pulse information is extracted for further 
localization. In the second step, single pulse cross spectrum is calibrated 
with phase reference source and then fitted to derive residual delay. 

The single pulse search and localization scheme described in 
Fig.~\ref{fig:solving} has been implemented as the ``VOLKS'' (VLBI Observation 
for single pulse Localization Keen Searcher) 
pipeline. At present,
this pipeline supports dedispersion, fringe fitting of fast dump cross 
spectrum, filtering of single pulse candidates from multiple re-sampling time 
(window length), multiple baselines cross matching. For localization, it 
supports both radio imaging and astrometric solving methods. The pipeline is 
still being improved, so as 
to support more features, e.g., cross spectrum based DM search, GPU 
acceleration 
of fringe fitting, etc. We have made this pipeline publicly 
available \citep[][Codebase: \url{https://github.com/liulei/volks}]{VOLKS}.

\section{Localization result}\label{sec:loc}
We carry out single pulse search and localization in a VLBI pulsar data set 
which has been used in \citet{Liu2018a}. All works except for the AIPS 
calibration 
part are carried out with the ``VOLKS'' pipeline described in 
Sec.~\ref{sec:intro_methods}. We present the localization results using both 
radio imaging and astrometric solving methods and compared their accuracies. 

\subsection{Data set}
\begin{figure*}
	\plotone{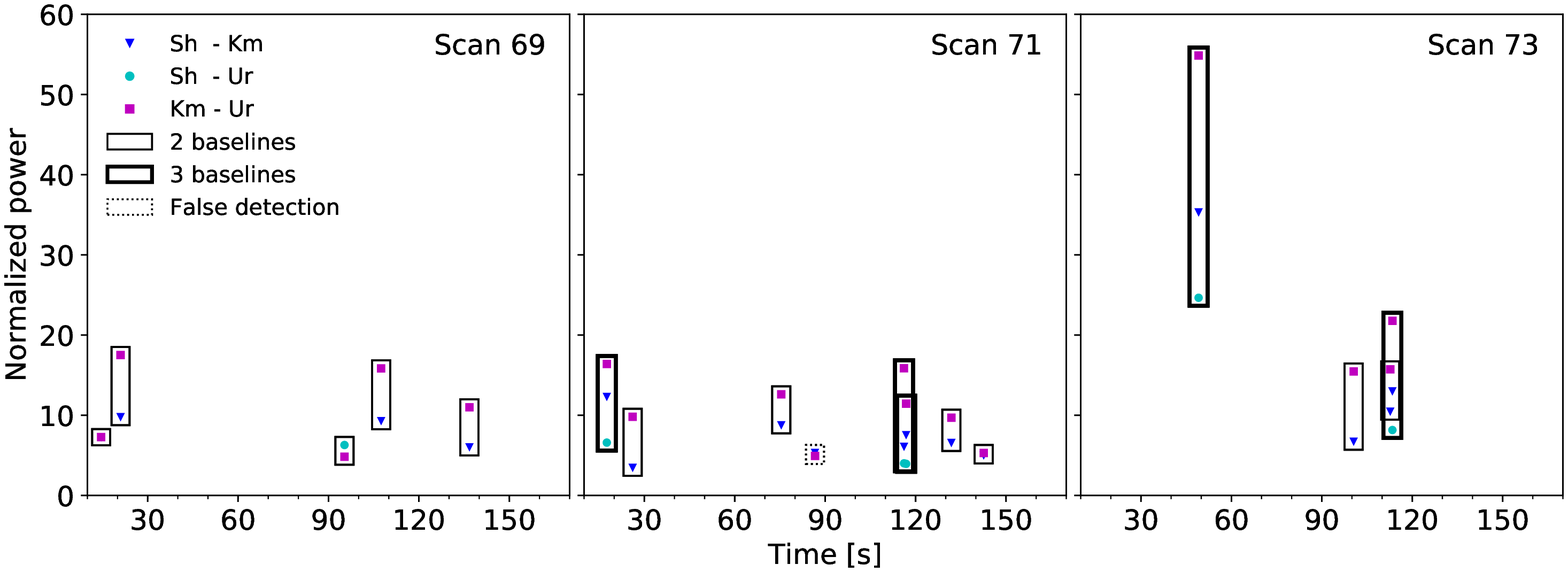}
	\caption{Single pulse detection result of VLBI pulsar data set psrf02. 
	Single pulses detected on multiple baselines are enclosed by rectangular 
	boxes. Thin and thick solid line rectangles correspond to those	detected on 
	2 and 3 
		baselines, respectively. The only one enclosed by dotted rectangle is 
		false 
		detection according to its pulsar phase. Note that the actual width of 
		the single pulse is much 
		narrower than the width of the rectangular box. In total, 17 single 
		pulses (including 1 false detection) are detected on 2 or 3 baselines. 
		The detailed parameters of these single pulses together with their 
		localization results are presented in Tab. \ref{tab:sp}. 
		\label{fig:cms}}
\end{figure*}

Data is taken from CVN VLBI pulsar observation psrf02. The 96 MHz 
bandwidth data in S band is recorded in 6 
frequency channels, 2 bits sampling. Three CVN stations, Sh, 
Km, Ur participate the observation. The details of the observation are
presented in \citet{Liu2018a}. Among the 293 scans in the 24 hours observation, 
single pulses of PSR J0332+5434 in Scan 69, 71 and 73 are extracted for 
localization. 
J0347+5557 in Scan 68, 70, 72 and 74 are used as phase reference source, 3C273 
in Scan 293 is used for PCAL, clock and channel delay calibration. 

Since CALC is easy to be integrated into the localization pipeline, in this 
work, we use CALC 9.1 for partial derivative, $uv$ and delay model 
calculation. In order to keep the consistency, we reprocess the raw data using 
DiFX 
correlator\footnote{DiFX use CALC for $uv$ and delay model calculation} and 
carry out single pulse search with exactly the same procedure described in 
\citet{Liu2018a}. 

Fig.~\ref{fig:cms} presents the single pulse detection result using DiFX 
correlator \citep{DiFX2007, DiFX2011}. We expect it to show identical result as 
Fig. 5 in \citet{Liu2018a} using the CVN software correlator \citep{Zheng2010}.
However, at first glance, they are not consistent with each other. 
According to our analysis, the main reason is, two correlators behave 
differently when SNR       {(Signal to Noise Ratio)} is low. In this case, when 
the 
normalized power is less than 5, the results are different. Since two 
correlators use totally different delay models, it is not 
surprising to give such kind of discrepancy. Besides that, the implementations 
of the 
algorithm in two correlators are different. 
The good thing is, when it comes to strong signals, the results given by two 
correlator are quite 
consistent: singles pulses at 17.5 s of Scan 71, 49.1 s and 113.4 s of Scan 
73 are detected on all 3 baselines by both correlators. However, the low 
sensitivity of Sh-Ur baseline still makes the result different: based on
DiFX output, two singles pulses at 116.2 s and 116.9 s of Scan 71 are detected 
on all 
three baselines, while the single pulse at 21.0 s of Scan 69 is missed on 
Sh-Ur baseline, although it is detected on all three baselines in 
\citet{Liu2018a}. In summary, 17 single pulses are detected on at least 2 
baselines. According to their pulsar phases, we may know the one enclosed 
by dotted rectangle is a false detection.

We extract the visibility records that contain single pulse information from 
the original visibility files, and convert them 
together with visibilities of phase reference source (J0347+5557) and 
calibration source (3C273) to FITS-IDI 
format\footnote{https://fits.gsfc.nasa.gov/registry/fitsidi.html} for further 
calibration and localization.

\subsection{Localization}

\begin{deluxetable*}{ccr|ccc|rr|rrr}
	\caption{Parameters of 17 single pulses detected in Scan 69, 71 and 73 of 
	CVN observation psrf02. All positions are given as offsets to a priori 
	position. The offset in Ra direction is the tangent plane projection: 
	$\Delta\alpha^*=\Delta\alpha\cos\delta$. For radio imaging, we take a 
	pixel size of 2.0 mas $\times$ 2.0 mas. SNR is 
	calculated as the peak flux subtracted by the average flux and then 
	normalized with noise (standard deviation). \label{tab:sp}}
	\tablewidth{0pt}
	\tablehead{\colhead{} & \colhead{} & \colhead{} & 
	\multicolumn{3}{c}{Baseline}&
	\multicolumn{2}{c}{Solving} & \multicolumn{3}{c}{Imaging} \\
	\colhead{No.} & \colhead{Scan} & \colhead{Time} & \colhead{Km-Sh} & 
	\colhead{Km-Ur} & \colhead{Sh-Ur} &
	\colhead{$\Delta\alpha^*$} & 
	\colhead{$\Delta\delta$} & \colhead{$\Delta\alpha^*$} & 
	\colhead{$\Delta\delta$} & \colhead{SNR} \\
		\colhead{} & \colhead{} & \colhead{(sec)} & \colhead{} & \colhead{} & 
		\colhead{} &
		\colhead{(mas)} & \colhead{(mas)} & \colhead{(mas)} & \colhead{(mas)} & 
		\colhead{}}
\startdata
 1 & 69 & 14.565 &  \checkmark & \checkmark &             & 421.9$\pm$17.9 & 
 -313.3$\pm$15.6 & 392.0 & -294.0 & 9.0 \\
2 & 69 & 21.000 &  \checkmark & \checkmark &             & 392.6$\pm$14.6 & 
-223.3$\pm$13.2 & 358.0 & -248.0 & 9.6 \\
3 & 69 & 95.306 &             & \checkmark & \checkmark  & 498.7$\pm$21.0 & 
-203.7$\pm$25.5 & 468.0 & -252.0 & 9.1 \\
4 & 69 & 107.463 &  \checkmark & \checkmark &             & 481.5$\pm$15.6 & 
-264.3$\pm$13.7 & 460.0 & -270.0 & 9.4 \\
5 & 69 & 136.765 &  \checkmark & \checkmark &             & 401.0$\pm$17.4 & 
-263.2$\pm$16.1 & 392.0 & -256.0 & 8.3 \\
6 & 71 & 17.547 &  \checkmark & \checkmark & \checkmark  & 377.9$\pm$10.2 & 
-327.2$\pm$13.4 & 350.0 & -306.0 & 11.1 \\
7 & 71 & 26.122 &  \checkmark & \checkmark &             & 579.7$\pm$18.1 & 
-187.6$\pm$15.9 & 518.0 & -222.0 & 8.9 \\
8 & 71 & 75.431 &  \checkmark & \checkmark &             & 444.9$\pm$16.0 & 
-216.3$\pm$15.1 & 410.0 & -192.0 & 8.4 \\
9 & 71 & 86.636 &  \checkmark & \checkmark &             & 7673.9$\pm$16.8 & 
-2873.4$\pm$15.4 & -26.0 & 540.0 & 6.5 \\
10 & 71 & 116.158 &  \checkmark & \checkmark & \checkmark  & 417.2$\pm$11.7 & 
-225.2$\pm$13.8 & 396.0 & -266.0 & 11.8 \\
11 & 71 & 116.872 &  \checkmark & \checkmark & \checkmark  & 389.6$\pm$11.7 & 
-290.3$\pm$15.8 & 358.0 & -288.0 & 10.1 \\
12 & 71 & 131.879 &  \checkmark & \checkmark &             & 416.5$\pm$16.4 & 
-228.2$\pm$15.1 & 382.0 & -234.0 & 9.1 \\
13 & 71 & 142.605 &  \checkmark & \checkmark &             & 414.4$\pm$17.5 & 
-301.3$\pm$16.3 & 422.0 & -290.0 & 8.8 \\
14 & 73 & 49.109 &  \checkmark & \checkmark & \checkmark  & 
386.2$\pm$\enspace8.2 & -280.1$\pm$\enspace9.4 & 388.0 & -284.0 & 10.7 \\
15 & 73 & 100.557 &  \checkmark & \checkmark &             & 447.5$\pm$17.3 & 
-190.4$\pm$15.5 & 432.0 & -204.0 & 7.7 \\
16 & 73 & 112.702 &  \checkmark & \checkmark &             & 442.7$\pm$15.1 & 
-245.5$\pm$14.6 & 416.0 & -240.0 & 8.6 \\
17 & 73 & 113.418 &  \checkmark & \checkmark & \checkmark  & 433.0$\pm$10.5 & 
-272.4$\pm$14.1 & 408.0 & -258.0 & 9.5 \\
\enddata
\end{deluxetable*}


\begin{figure}
	\plotone{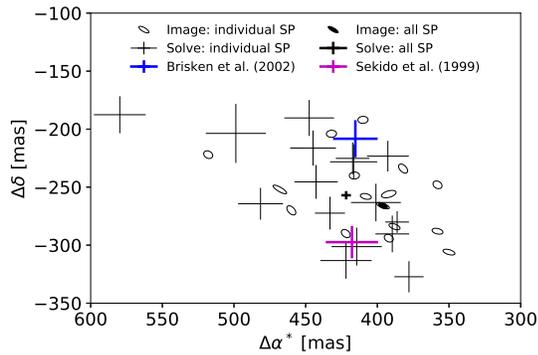}
	\caption{Localization of every individual single pulse (SP) and the average 
		result. For comparison, two reference positions are also presented. 
		They are calculated by evolving positions at reference epoch to the 
		date of pulsar observation according to given proper motions. Error 
		bars of reference positions are 
		calculated by combing uncertainties of positions and proper motions: 
		$\sigma=\sqrt{\sigma_\mathrm{pos}^2+(\sigma_\mathrm{pm}\Delta 
		t)^2}$.	Here $\sigma_\mathrm{pos}$ and $\sigma_\mathrm{pm}$ are  
		uncertainties of position and proper motions in Ra and Dec directions 
		given by references. $\Delta t$ is the time between reference epoch 
		and date of pulsar observation. Ellipses are drawn according to the 
		shape and position angle of the corresponding beams, which are 
		estimated using the same algorithm as in DIFMAP (TJP's algorithm). 
		All positions are given as offset to a 
		priori position. The offset in Ra direction is the tangent plane 
		projection: $\Delta\alpha^*=\Delta\alpha\cos\delta$. 
		\label{fig:loc_ref}}
\end{figure}

\begin{deluxetable*}{cccccc}[t]
	\caption{Reference positions (J2000.0) and proper motions of PSR J0332+5434
	at their respect reference epochs. The proper motion in Ra
	direction is the tangent plane projection: 
	$\mu^*_\alpha=\mu_\alpha\cos\delta$. \label{tab:ref_pos}}
	\tablewidth{0pt}
	\tablehead{\colhead{Reference} & \colhead{$\alpha$} & \colhead{$\delta$} & 
	\colhead{$\mu^*_\alpha$} & \colhead{$\mu_\delta$} & \colhead{Reference} \\
	\colhead{} & \colhead{} & \colhead{} & \colhead{(mas/yr)} & 
	\colhead{(mas/yr)} &
	\colhead{epoch}}
	\startdata
	\citet{Sekido1999} & $03^\mathrm{h}32^\mathrm{m}59^{s}.3760\pm0.0010$ &
	$54^\circ34^\prime43^{\prime\prime}.5040\pm0.0070$ & 17.30$\pm$0.80 &
	-11.50$\pm$0.60 & 1995.0\\
	\citet{Brisken2002} & $03^\mathrm{h}32^\mathrm{m}59^{s}.3862\pm0.0017$ &
	$54^\circ34^\prime43^{\prime\prime}.5051\pm0.0150$ & 17.00$\pm$0.27 &
	-9.48$\pm$0.37 & 2000.0\\
	\enddata
\end{deluxetable*}

We use AIPS (31DEC18) for calibration.  According to the
standard recipe for phase reference observations, the whole process 
consists of 3 steps: (a) Calibrate delay and phase in every individual 
frequency channel (IF) for single pulses and phase reference source 
(J0347+5557) using the solution derived from calibration source (3C273). (b) 
Derive solutions for phase reference source, including delays (combining all 
IFs), phases and delay rates, then interpolate them to the pulsar scans. (c) 
Calibrate every single pulse using solutions interpolated from phase reference 
source and output them with FITS-IDI formats for further localization. 
One thing we want to point out is, the above calibration procedure is only 
intended for our testing pulsar data set. In real FRB search, the burst and the 
target source appear in the same FoV. The corresponding calibration is somewhat 
similar with phase reference observation, but more simplified. Please refer to
Sec.~\ref{sec:search_geodetic} for a detailed explanation.

Our implementation of radio imaging takes similar procedure as in DIFMAP 
\citep{DIFMAP}. The main difference 
is DIFMAP only deals with data that all frequency points inside one IF are 
summed together, which greatly reduces the time consumption of imaging process
at the expense of small imaging area. However,
this is based on the assumption that target source is close to its 
a priori position and therefore no fringe phase ambiguity exists inside one IF. 
For single pulse search, large ambiguities might still exist after phase 
calibration if the burst is far from a priori position. To keep 
full fringe phase ambiguity information, in this work, frequency points inside 
one IF are not summed together before they are gridded in the $uv$ plane. 

To calculate the delay model of PSR J0332+5434 for VLBI correlation, we use a 
priori position (Ra: 3$^\mathrm{h}$32$^\mathrm{m}$59$^\mathrm{s}$.368, Dec: 
54$^\circ$34$^\prime$43$^{\prime\prime}$.57, in J2000.0) given by ATNF pulsar 
database\footnote{http://www.atnf.csiro.au/research/pulsar/psrcat} \citep{ATNF} 
at reference epoch MJD 46473 (Feb. 12, 1986). The localization results are 
presented in Tab.~\ref{tab:sp}. According to uncertainties of each single pulse 
given by astrometric solving, the results derived by both methods are 
consistent 
with each other in a 3$\sigma$ level. Compared with 2 baselines results, 3 
baselines results usually yield smaller uncertainties and higher SNR. Among all 
single pulses presented here, No. 14 corresponds to the one with the highest 
normalized power in Fig.~\ref{fig:cms} and the smallest 
uncertainties. Note that its SNR is not the highest, according to our 
investigation, this is due to its large flux 
density 
fluctuation in image plane.
Also note No. 9 in the table, which corresponds to the false detection in 
Fig.~\ref{fig:cms}. Clearly it yields incorrect position with both radio 
imaging and astrometric solving methods. Moreover, positions given by two 
methods are inconsistent with each other. 


By griding 
all single pulses (except for the false detection) in the $uv$ plane, we obtain 
the 
average position offset ($\Delta\alpha^*$: 396~mas, $\Delta\delta$: -266~mas). 
By solving linear equations of all single pulses (except the false detection) 
together, we derive the average position correction ($\Delta\alpha^*$: 
421.7$\pm$3.3~mas, $\Delta\delta$: -257.0$\pm$3.6~mas). Mathematically, the 
two methods are equivalent. We expect them to give identical results. However, 
in the actual data processing, the imaging and fringe phase fitting procedures
are totally different. For instance, the weight of each single pulse is 
determined by 
the amplitude of cross spectrum and the scatter of fringe phase, respectively.
This leads to the discrepancy of the average position. 

In Fig.~\ref{fig:loc_ref}, we plot the localization result of every individual 
single pulse and average positions given by two methods. To evaluate the 
absolute localization 
precision of the whole data processing pipeline, we also present two reference
positions \citep[][see Tab.~\ref{tab:ref_pos} for details]{Sekido1999, 
Brisken2002}. The two positions are 
derived by evolving the reference positions at reference epochs to pulsar 
observation date according to reference proper motions. As demonstrated in the 
figure, reference positions are roughly consistent with single pulse 
localization result. All single pulses except one 
derived by astrometric solving method distribute in a 200 mas $\times$ 200 mas 
area. The scatters\footnote{The scatter is estimated by combining the standard 
deviations in Ra and Dec directions: $\sigma = \sqrt{\sigma_{\Delta\alpha^*}^2 
+ \sigma_{\Delta\delta}^2}$.}
 of single pulses locations derived by radio imaging and 
astrometric solving methods are 53.2 mas and 65.1 mas, respectively, which can 
be regarded as the absolute localization precision of this work. 

\section{Analysis of computational cost}\label{sec:computation}
\begin{figure}
	\plotone{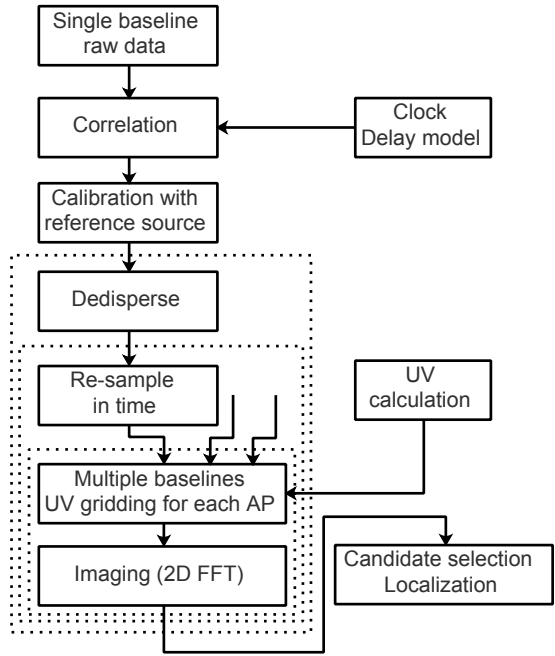}
	\caption{Demonstration of single pulse search with radio imaging pipeline. 
		Steps enclosed by dotted rectangles represent iterations with 
		corresponding quantities.\label{fig:imaging}}
\end{figure}

     {In this section, we analyze the computational cost of both radio 
imaging and geodetic VLBI based single pulse search pipeline, and 
come to the conclusion that the letter one is more suitable for FRB search in 
real observation.}

Fig.~\ref{fig:imaging} demonstrates the single pulse search with 
radio imaging pipeline. Visibilities from each baseline are first calibrated 
and 
then transformed to the image plane to create fast dumped images with multiple 
re-sampling times. Single pulse candidates are detected in these fast dumped 
images according to given threshold. To use FFT to speedup the transformation 
process, visibilities are gridded in the $uv$ plane \citep{Thompson2001}. 
In radio imaging search pipeline, the search and localization steps are coupled 
together: single pulses are detected and localized directly in the fast 
dumped images. However, such kind of scheme is only feasible for VLBI system 
with not very long baselines. Taking the configuration of VLA for example. The 
longest baseline is 36 km. In S band (2.2 GHz), the angular resolution is 
around 0.78 arcsec. The diameter of VLA antenna is 25 m. As a raw estimation, 
the corresponding FoV is 22.88 arcmin      {($\sim1.22~\lambda/D$)}. In radio 
imaging, the pixel size usually takes a quarter of angular resolution.      {To 
cover 
80\% of the FoV ($\sim\lambda/D$), the corresponding map 
size is 5760$\times$5760, which is reasonable for 2D FFT and the single 
pulse detection afterwards. However, for 
a typical VLBI network, e.g., Chinese VLBI Network \citep[CVN,][]{CVN}, the 
baseline is as long as 3000 km. By keeping other parameters 
unchanged, the corresponding map size is      {480000$\times$480000}, which is 
two orders of magnitude larger than that of VLA.} Since the computational 
complexity 
of radio imaging pipeline is usually proportional to map size, the 
computational cost is two orders of magnitude higher. Obviously this is 
a huge challenge for the actual operation.

It is possible to compare the computational cost of both pipelines 
in an qualitative way: by investigating Fig.~\ref{fig:solving} and 
Fig.~\ref{fig:imaging}, one may find that both pipelines involve three 
loops. From outside to inside, they are: DM 
trials for dedispersion, multiple re-sampling times and series of re-sampled 
cross spectrum. By selecting the same re-sampling time, 
the two pipelines require the same number of iterations.      {For radio 
imaging, the inner most operation involves $uv$ gridding, 2D FFT and 
single pulse selection. The first term is negligible as it is proportional to 
the number of samples. In contrast, in the geodetic VLBI pipeline, the 
most time consuming part is fringe fitting, which involves 2D FFT to search for 
single band delay (SBD) and multi band delay (MBD). We may demonstrate that the 
size of this 2D array is much smaller than that of radio imaging. E.g., for 
a typical configuration of the pulsar observation in the above section, 
Tab.~\ref{tab:complexity} present the computational complexity of both 
pipelines. For geodetic VLBI pipeline, according to Eq. 12 of~\citet{Liu2018a}, 
the 
minimum FFT size for SBD search is 1746, the number of 
frequency channels for MBD search is 6. By rounding them to the power of 2 and 
taking 4 times extrapolation, the corresponding sizes are 8192 and 32, 
respectively. 
One may find that for a baseline length of 3000 km in CVN, the actual 
computational cost of radio imaging is much higher than that of geodetic 
VLBI pipeline. It is not difficult to come to the conclusion that 
the geodetic VLBI pipeline is more suitable for real time FRB 
search than radio imaging pipeline.}
 
\begin{deluxetable*}{ll|ll}
	\caption{Computational complexity of two single pulse search pipelines. 
	\label{tab:complexity}}
	\tablewidth{0pt}
	\tablehead{
		\multicolumn{2}{c}{Radio imaging} & \multicolumn{2}{c}{Geodetic VLBI}\\
		\multicolumn{2}{c}{N: 480000}	& \multicolumn{2}{c}{$N_1$: 8192, 
		$N_2$: 32} \\
	}
	\startdata
	2D FFT	&	$2\times\frac{34}{9}N\log_2(N)\times N$ & 2D FFT (per baseline) 
	& 
	$\frac{34}{9}N_1\log_2 (N_1) \times N2 + \frac{34}{9}N_2\log_2 (N_2) \times 
	N_1$ \\
	Finding peak & $N^2$	& Finding peak (per baseline) & $N_1\times N_2$ \\
	Total	& $\frac{68}{9}N^2\log_2(N) + N^2$ & Total (3 baselines) & 
	$\frac{34}{3}N_1 N_2\log_2(N_1 N_2) + 3 N_1 N_2$ \\
	& ($3.31\times10^{13}$) & & ($5.43\times10^7$) \\
	\enddata
\end{deluxetable*}

\section{Discussion}\label{sec:discuss}
\subsection{Subtraction of constant sources}
Single pulse search is usually carried out as commensal task in regular VLBI 
observations. In this case, how to remove the influence of target source and 
other constant sources is a problem that must be solved. The 
``realfast'' pipeline deals with this problem by subtracting the mean 
visibility 
in time on timescales less than the VLA fringe rate. Similar treatment is 
suitable for our geodetic VLBI based pipeline, too. Besides that, we propose 
another scheme which is specially designed for fringe fitting pipeline: to 
carry out single pulse search, the clock is well adjusted (fringe rate less 
than $10^{-3}$~Hz), the MBD and SBD for the target source does not change too 
much in the whole scan. We may skip the target source and the surrounding area 
in the MBD and SBD search matrix. However, this scheme can only be verified 
with data in which fast transients present together with constant source, which 
is not available at present. In this case, mean visibilities subtraction scheme 
might be more reasonable.

\subsection{Large search area}\label{sec:large_fov}
One might doubt that if it is possible to detect single pulses
efficiently in the whole FoV with our method. In \citet{Liu2018a}, we point out 
that the fringe fitting process is somewhat similar with that of coherent beam 
forming, but without the computational expense to form a great number of beams 
to cover the whole FoV of telescopes. For localization, traditionally, we only 
carry out narrow field imaging\footnote{By assuming the burst is a point 
source, mathematically radio imaging and astrometric solving methods are 	
equivalent.}. The corresponding searching area is much smaller than the 	
FoV. The main reason is, the delay model is (usually) calculated for the 	
center of FoV but applied to the whole FoV. The residual delay rate is large at 
the edge of the FoV. The signal degrades quickly as the integration becomes 
long. In regular VLBI correlation, the integration time is as long as 1 
second. In contrast, in our single pulse detection method, the maximum 
integration time is no more than 32 ms, and is usually as short as 4~ms. This 
short integration time makes it possible to investigate the whole FoV with only 
one image. Actually this is somewhat similar with the implementation of 
multiple phase center in modern VLBI correlator \citep{DiFX2011, SFXC}. For 
instance, in SFXC, for each sub-integration period (25 ms), a phase shift is 
performed for each phase center, so as to compensate for the phase change due 
to the large residual delay rate in that position. We know that although the 
signal will not degrade very much within such a short time, the corresponding 
SNR is low. In this work, we have demonstrated that it is still 
possible to detect signals for such a short integration time.

For the localization of single pulse in the whole FoV, one of the drawbacks of 
radio imaging is the computational cost increases significantly when the 
imaging area becomes large in long baseline observation. 
In contrast, in the geodetic VLBI search scheme, this is not a problem. We 
have to admit 
that when the single pulse is far from phase center, e.g., close to 
$\frac{1}{2}~\theta_\mathrm{FWHM}$, the 
performance of the search pipeline is still not clear. Possible problems 
include 
the decrease of detection sensitivity, the increase of localization 
uncertainty, etc. Although these problems 
also exist in the radio imaging based search scheme, we have not seen their 
related descriptions and solutions. 
Therefore, we propose to carry out further VLBI observation to test the 
pipeline. 
For instance, by placing the pulsar in the FoV with different offsets to the 
phase 
center, such that we may plot the power and localization precision of the 
detected single pulses as a function of offset to FoV center.  

\subsection{Localization in geodetic VLBI observation} 
\label{sec:search_geodetic}
     {The VLBI observation that provides the pulsar data set used in this work 
is 
carried out in phase reference mode, which makes it possible to calibrate the 
extracted single pulses with phase reference source. However, geodetic VLBI 
obervation takes a totally different approach: to cover as large sky area as 
possible, sources distribute evenly in the sky. In this case, it is still 
possible to localize the burst: the target source itself is a perfect phase 
reference source. Since the burst and the target source always appear in 
the same FoV, it is even not necessary to extrapolate the fringe fitting 
solution 
to the burst time. We may just derive MBD, SBD, delay rate and residual phase 
for the target source in the scan, and then calibrate the visibilities with 
these quantities. Single pulse search is carried out with calibrated 
visibilities. Once single pulse is detected, the derived delay can be used 
directly for localization with astrometric method. No further calibration with 
phase reference source is needed.}

\subsection{Dispersion measure search}\label{sec:dm_search}
      {One thing we want to point out is we do not carry out dispersion 
measure search
in this work. The DM value provided by ATNF pulsar database is used for 
dedispersion.}
The main reason is the DM value of PSR J0332+5434 is too low 
(26.833 $\mathrm{pc~cm}^{-3}$). 
In \citet{Liu2018a}, we have proposed a DM search scheme, which is quite 
straight forward: dividing the DM search range into multiple bins, then 
carrying out dedispersion and single pulses search independently for these 
bins. For a minimum re-sampling 
time of 4.096~ms and a frequency range of 2192~MHz to 2288~MHz, the 
corresponding DM resolution is 57.7 $\mathrm{pc~cm}^{-3}$, 
which is too large to resolve the DM of this pulsar. Since DM search is an 
important part in the whole single pulse 
search pipeline, we propose to observe RRAT \citep{RRAT} sources to obtain high 
DM data to test our geodetic VLBI based search pipeline. Our observation 
proposal has been submitted to EVN, and is scheduled in March, 2019.

\section{Conclusions}\label{sec:conclusion}
     {In this paper, we present the astrometric solving based single pulse 
localization method. By applying this method to a VLBI pulsar observation data 
set, we demonstrate that the localization result for each single pulse derived 
by both radio imaging and astrometric solving are consistent with each other in 
a 3 $\sigma$ level. Most of single pulses, together with reference positions,
distribute in a 200~mas $\times$ 200~mas area. The scatters of localization 
results using both methods are less than 70~mas, which can be regarded as 
the absolute localization precision. Our work proves that it is possible to 
derive single pulse position with reasonable precision based on just 3 or even 
2 baselines and 4 ms integration in VLBI observation.
The localization method, together with the single pulse search method in 
\citet{Liu2018a},  build up the complete geodetic VLBI based single pulse 
search and localization pipeline. We further demonstrate that the computational 
cost of radio imaging pipeline is much higher than that of geodetic VLBI 
based pipeline. Therefore, for cross spectrum based FRB search in VLBI 
observation, geodetic VLBI pipeline might be a better choice.}
We name our pipeline as ``VOLKS'' and have made it publicly available. We hope 
this will be helpful for radio transient studies.

\acknowledgments
We thank Fengxian Tong for helpful discussions on VLBI point source 
localization. This work is supported by the Joint Research Fund in Astronomy 
(U1631122, U1831137) under cooperative agreement between the National Natural 
Science 
Foundation of China (NSFC) and Chinese Academy of Sciences (CAS), the 
Natural Science Foundation of China (11573057, 11403073, 11633007, 11703070), 
the Shanghai Key Laboratory of Space Navigation and Positioning Techniques 
(15DZ2271700), the Ten Thousand Talent Program, the CAS Key Technology Talent 
Program, Shanghai Outstanding Academic Leaders Plan, the strategic 
Priority Research Program of Chinese Academy of Sciences, Grant No. XDB23010200 
and the Knowledge Innovation Program of the Chinese Academy of Sciences (Grant 
No. KJCX1-YW-18). W.L thanks the support of World  Premier  International
Research   Center   Initiative   (WPI   Initiative), Japan. 

\appendix
\section{Least square solutions}\label{append:lsq}
The derive of offset ($\Delta\alpha, \Delta\delta$) to a priori position is 
divided into two steps. 

(a) Fitting residual delay. For every single pulse, the delay $\tau$ for each 
baseline is derived by fitting the fringe phase $\phi_k$ after phase reference 
calibration as a function of frequency $f_k$:
\begin{equation}
\phi_k = 2\pi f_k\tau + \phi_0.
\end{equation}
The fit of above linear equation by using the amplitude of each frequency point 
$f_k$ as weight is available in most mathematical libraries. After fitting, we 
obtain the delay $\tau_i$ and the corresponding uncertainties $\sigma_i$ for 
baseline $i$. The relation between $\sigma_i$ and the scatter of fringe phase 
is 
explained in \citet{Takahashi2000}. Note that when the source is far from a 
priori position, fringe phase ambiguity exists even after calibration. In 
this case we have to compensate an initial delay value to remove ambiguity 
before fitting\footnote{When the scatter of fringe phase is large, simply 
unwarp the fringe phase and then do linear fitting might lead to incorrect 
result.}. 

(b) Derive position offset. This is to solve the linear equation:
\begin{equation}
\mathbf{y}=\mathbf{A}\mathbf{x}.
\end{equation}
Here $\mathbf{y}=(\tau_1, \tau_2, ..., \tau_n)^T$ is the delay vector for $n$ 
baselines. $\mathbf{x}=(\Delta\alpha, \Delta\delta)^T$ is the position offset 
vector. 
$\mathbf{A}=(A_1, A_2, ..., A_n)^T$ is the partial derivative matrix: $A_i = 
\left(\frac{\partial\tau_i}{\partial\alpha}, 
\frac{\partial\tau_i}{\partial\delta}\right)$.
The least square solution of above equations is:
\begin{equation}
\mathbf{\hat{x}}=(\mathbf{A}^T\mathbf{WA})^{-1}\mathbf{A}^T\mathbf{Wy}.
\end{equation}
Here $\mathbf{W}$ is the weight matrix: $\mathbf{W}=\mathbf{\Sigma}^{-1}$. 
$\mathbf{\Sigma}=\mathrm{diag}(\sigma_1^2, \sigma_2^2, ..., \sigma_n^2)$ is the 
delay error matrix. The estimation parameter error matrix is: 
\begin{equation}
\mathbf{\Sigma_p}=(\mathbf{A}^T\mathbf{\Sigma}^{-1}\mathbf{A})^{-1}.
\end{equation} 

The square root of $\Sigma_{\mathrm{p},11}$ and $\Sigma_{\mathrm{p},22}$  
correspond to uncertainties of $\Delta\alpha$ and $\Delta\delta$, respectively.

\end{document}